\documentclass[%
 reprint,
superscriptaddress,
showpacs,
 amsmath,amssymb,
 aps, prd
]{revtex4-1}

\usepackage{graphicx}
\usepackage{dcolumn}
\usepackage{bm}
\usepackage{hyperref}
\usepackage{subfigure}
\usepackage{comment}
\usepackage{xcolor}
\usepackage[mathlines]{lineno}

\usepackage{ifthen} 
\newboolean{pdflatex}
\setboolean{pdflatex}{true} 

\newboolean{articletitles}
\setboolean{articletitles}{true} 

\newboolean{uprightparticles}
\setboolean{uprightparticles}{false} 

\newboolean{inbibliography}
\setboolean{inbibliography}{false} 

\newboolean{wordcount}
\setboolean{wordcount}{false} 

\ifthenelse{\boolean{wordcount}}%
{\usepackage{bibentry} 
 \usepackage{mciteplus} 
 \usepackage{comment} 
 \excludecomment{acknowledgments} 
 \def\maketitle{} 
}{
}

\graphicspath{{./figs/}}
\DeclareGraphicsExtensions{.pdf,.PDF,png,.PNG}
\usepackage{xspace} 
\makeatletter
\gdef\@ptsize{0} 
\makeatother
\makeatletter
\usepackage{suffix}
\let\latex@section\section
\WithSuffix\def\section*{\secdef\my@section{\latex@section*}}
\def\my@section[#1]#2{}
\makeatother

\usepackage{xspace} 
\usepackage{upgreek}

\usepackage{relsize}
\def\babar{\mbox{\slshape B\kern-0.1em{\smaller A}\kern-0.1em
    B\kern-0.1em{\smaller A\kern-0.2em R}}\xspace}

\def\MagUp {\mbox{\em Mag\kern -0.05em Up}\xspace}

\ifthenelse{\boolean{uprightparticles}}%
{
 
 \def\Pgamma      {\ensuremath{\upgamma}\xspace}

 \def\Peta        {\ensuremath{\upeta}\xspace}

 \def\Pmu         {\ensuremath{\upmu}\xspace}

 \def\Ppi         {\ensuremath{\uppi}\xspace}

 \def\Ptau        {\ensuremath{\uptau}\xspace}

 \def\Ppsi        {\ensuremath{\uppsi}\xspace}

 \def\PDelta      {\ensuremath{\Delta}\xspace}                 
 \def\PXi      {\ensuremath{\Xi}\xspace}                 
 \def\PLambda      {\ensuremath{\Lambda}\xspace}                 
 \def\PSigma      {\ensuremath{\Sigma}\xspace}                 
 \def\POmega      {\ensuremath{\Omega}\xspace}                 
 \def\PUpsilon      {\ensuremath{\Upsilon}\xspace}                 
 
 \def\PB      {\ensuremath{\mathrm{B}}\xspace}                 
                  
 \def\PD      {\ensuremath{\mathrm{D}}\xspace}

 \def\PJ      {\ensuremath{\mathrm{J}}\xspace}                 
 \def\PK      {\ensuremath{\mathrm{K}}\xspace}

 \def\Pb      {\ensuremath{\mathrm{b}}\xspace}

 \def\Pe      {\ensuremath{\mathrm{e}}\xspace}

 \def\Pi      {\ensuremath{\mathrm{i}}\xspace}

}
{
 
 \def\Pgamma      {\ensuremath{\gamma}\xspace}

 \def\Peta        {\ensuremath{\eta}\xspace}

 \def\Pmu         {\ensuremath{\mu}\xspace}

 \def\Ppi         {\ensuremath{\pi}\xspace}

 \def\Ptau        {\ensuremath{\tau}\xspace}

 \def\Ppsi        {\ensuremath{\psi}\xspace}                 
                  
 \mathchardef\PDelta="7101
 \mathchardef\PXi="7104
 \mathchardef\PLambda="7103
 \mathchardef\PSigma="7106
 \mathchardef\POmega="710A
 \mathchardef\PUpsilon="7107
                  
 \def\PB      {\ensuremath{B}\xspace}                 
                  
 \def\PD      {\ensuremath{D}\xspace}

 \def\PJ      {\ensuremath{J}\xspace}                 
 \def\PK      {\ensuremath{K}\xspace}

 \def\Pb      {\ensuremath{b}\xspace}

 \def\Pe      {\ensuremath{e}\xspace}

 \def\Pi      {\ensuremath{i}\xspace}

}

\makeatletter
\ifcase \@ptsize \relax
  \newcommand{\miniscule}{\@setfontsize\miniscule{4}{5}}
\or
  \newcommand{\miniscule}{\@setfontsize\miniscule{5}{6}}
\or
  \newcommand{\miniscule}{\@setfontsize\miniscule{5}{6}}
\fi
\makeatother

\DeclareRobustCommand{\optbar}[1]{\shortstack{{\miniscule (\rule[.5ex]{1.25em}{.18mm})}
  \\ [-.7ex] $#1$}}

\def\en         {{\ensuremath{\Pe^-}}\xspace}   
\def\ep         {{\ensuremath{\Pe^+}}\xspace}

\def\mup        {{\ensuremath{\Pmu^+}}\xspace}
\def\mun        {{\ensuremath{\Pmu^-}}\xspace} 

\def\mumu       {{\ensuremath{\Pmu^+\Pmu^-}}\xspace}

\def\taup       {{\ensuremath{\Ptau^+}}\xspace}
\def\taum       {{\ensuremath{\Ptau^-}}\xspace}

\def\ellm       {{\ensuremath{\ell^-}}\xspace}
\def\ellp       {{\ensuremath{\ell^+}}\xspace}
\def\ellell     {\ensuremath{\ell^+ \ell^-}\xspace}

\def\g      {{\ensuremath{\Pgamma}}\xspace}

\def\bquark    {{\ensuremath{\Pb}}\xspace}

\def\pion   {{\ensuremath{\Ppi}}\xspace}
\def\piz    {{\ensuremath{\pion^0}}\xspace}
\def\pip    {{\ensuremath{\pion^+}}\xspace}
\def\pim    {{\ensuremath{\pion^-}}\xspace}

\def\pimp   {{\ensuremath{\pion^\mp}}\xspace}

\def\kaon    {{\ensuremath{\PK}}\xspace}
  \def\Kbar    {{\kern 0.2em\overline{\kern -0.2em \PK}{}}\xspace}

\def\KorKbar    {\kern 0.18em\optbar{\kern -0.18em K}{}\xspace}
\def\Kz      {{\ensuremath{\kaon^0}}\xspace}
\def\Kzb     {{\ensuremath{\Kbar{}^0}}\xspace}
\def\Kp      {{\ensuremath{\kaon^+}}\xspace}
\def\Km      {{\ensuremath{\kaon^-}}\xspace}
\def\Kpm     {{\ensuremath{\kaon^\pm}}\xspace}

\def\KS      {{\ensuremath{\kaon^0_{\mathrm{ \scriptscriptstyle S}}}}\xspace}
\def\KL      {{\ensuremath{\kaon^0_{\mathrm{ \scriptscriptstyle L}}}}\xspace}
\def\Kstarz  {{\ensuremath{\kaon^{*0}}}\xspace}

\def\Kstar   {{\ensuremath{\kaon^*}}\xspace}

\def\Kstarp  {{\ensuremath{\kaon^{*+}}}\xspace}

\newcommand{\etapr}{\ensuremath{\Peta^{\prime}}\xspace}

  \def\Dbar    {{\kern 0.2em\overline{\kern -0.2em \PD}{}}\xspace}
\def\D       {{\ensuremath{\PD}}\xspace}

\def\DorDbar    {\kern 0.18em\optbar{\kern -0.18em D}{}\xspace}
\def\Dz      {{\ensuremath{\D^0}}\xspace}
\def\Dzb     {{\ensuremath{\Dbar{}^0}}\xspace}
\def\Dp      {{\ensuremath{\D^+}}\xspace}
\def\Dm      {{\ensuremath{\D^-}}\xspace}

\def\Dstar   {{\ensuremath{\D^*}}\xspace}

\def\Dstarz  {{\ensuremath{\D^{*0}}}\xspace}

\def\Dstarp  {{\ensuremath{\D^{*+}}}\xspace}
\def\Dstarm  {{\ensuremath{\D^{*-}}}\xspace}

\def\B       {{\ensuremath{\PB}}\xspace}
\def\Bbar    {{\ensuremath{\kern 0.18em\overline{\kern -0.18em \PB}{}}}\xspace}

\def\BorBbar    {\kern 0.18em\optbar{\kern -0.18em B}{}\xspace}
\def\Bz      {{\ensuremath{\B^0}}\xspace}
\def\Bzb     {{\ensuremath{\Bbar{}^0}}\xspace}
\def\Bu      {{\ensuremath{\B^+}}\xspace}
\def\Bub     {{\ensuremath{\B^-}}\xspace}

\def\Bpm     {{\ensuremath{\B^\pm}}\xspace}

\def\jpsi     {{\ensuremath{{\PJ\mskip -3mu/\mskip -2mu\Ppsi\mskip 2mu}}}\xspace}

  \def\Y#1S{\ensuremath{\PUpsilon{(#1S)}}\xspace}

\def\FourS {{\Y4S}}

\def\Deltares    {{\ensuremath{\PDelta}}\xspace}

\def\Lz          {{\ensuremath{\PLambda}}\xspace}
\def\Lbar        {{\ensuremath{\kern 0.1em\overline{\kern -0.1em\PLambda}}}\xspace}
\def\LorLbar    {\kern 0.18em\optbar{\kern -0.18em \PLambda}{}\xspace}

\def\Lb      {{\ensuremath{\Lz^0_\bquark}}\xspace}
\def\Lbbar   {{\ensuremath{\Lbar{}^0_\bquark}}\xspace}

\def\BF         {{\ensuremath{\mathcal{B}}}\xspace}

\def\BR         {\BF}

\def\to                 {\ensuremath{\rightarrow}\xspace}

\def\qsq       {{\ensuremath{q^2}}\xspace}

\def\CP                {{\ensuremath{C\!P}}\xspace}

\def\AT#1     {\ensuremath{A_{\mathrm{T}}^{#1}}\xspace}           

\def\ctl       {\ensuremath{\cos{\theta_\ell}}\xspace}

\def\C#1      {\ensuremath{\mathcal{C}_{#1}}\xspace}                       
\def\Cp#1     {\ensuremath{\mathcal{C}_{#1}^{'}}\xspace}                    
\def\Ceff#1   {\ensuremath{\mathcal{C}_{#1}^{\mathrm{(eff)}}}\xspace}        
\def\Cpeff#1  {\ensuremath{\mathcal{C}_{#1}^{'\mathrm{(eff)}}}\xspace}       
\def\Ope#1    {\ensuremath{\mathcal{O}_{#1}}\xspace}                       
\def\Opep#1   {\ensuremath{\mathcal{O}_{#1}^{'}}\xspace}                    

\newcommand{\tev}{\ifthenelse{\boolean{inbibliography}}{\ensuremath{~T\kern -0.05em eV}}{\ensuremath{\mathrm{\,Te\kern -0.1em V}}}\xspace}
\newcommand{\gev}{\ensuremath{\mathrm{\,Ge\kern -0.1em V}}\xspace}
\newcommand{\mev}{\ensuremath{\mathrm{\,Me\kern -0.1em V}}\xspace}
\newcommand{\kev}{\ensuremath{\mathrm{\,ke\kern -0.1em V}}\xspace}
\newcommand{\ev}{\ensuremath{\mathrm{\,e\kern -0.1em V}}\xspace}
\newcommand{\gevc}{\ensuremath{{\mathrm{\,Ge\kern -0.1em V\!/}c}}\xspace}
\newcommand{\mevc}{\ensuremath{{\mathrm{\,Me\kern -0.1em V\!/}c}}\xspace}
\newcommand{\gevcc}{\ensuremath{{\mathrm{\,Ge\kern -0.1em V\!/}c^2}}\xspace}
\newcommand{\gevgevcccc}{\ensuremath{{\mathrm{\,Ge\kern -0.1em V^2\!/}c^4}}\xspace}
\newcommand{\mevcc}{\ensuremath{{\mathrm{\,Me\kern -0.1em V\!/}c^2}}\xspace}

\def\mus  {\ensuremath{{\,\upmu{\mathrm{ s}}}}\xspace}

\def\gsim{{~\raise.15em\hbox{$>$}\kern-.85em
          \lower.35em\hbox{$\sim$}~}\xspace}
\def\lsim{{~\raise.15em\hbox{$<$}\kern-.85em
          \lower.35em\hbox{$\sim$}~}\xspace}

\def\degrees{\ensuremath{^{\circ}}\xspace}

\def\tell1  {TELL1\xspace}
\def\ukl1   {UKL1\xspace}

\def\Ltophll      {\ensuremath{\Lb \to p h^- \ellm \ellp}\xspace}
\def\LJLz      {\ensuremath{\Lb \to \Lz \jpsi }\xspace}
\def\LJLzp     {\ensuremath{\Lb \to \Lz(\to p \pim) \jpsi(\to \mun \mup) }\xspace}

\def\dzz       {\ensuremath{d^2_{00}}}
\def\thetal    {\ensuremath{\theta_\ell}}
\def\phil      {\ensuremath{\phi_\ell}}
\def\thetah    {\ensuremath{\theta_h}}
\def\phih      {\ensuremath{\phi_h}}

\def\cth       {\ensuremath{\cos{\theta_h}}\xspace}

\def\pbee     {\ensuremath{P_b}\xspace}
\def\allam     {\ensuremath{\alpha_\Lambda}\xspace}
\def\apsq     {\ensuremath{|A^L_+|^2}\xspace}
\def\amsq     {\ensuremath{|A^L_-|^2}\xspace}
\def\bpsq     {\ensuremath{|B^L_+|^2}\xspace}
\def\bmsq     {\ensuremath{|B^L_-|^2}\xspace}

\newcommand {\img} {\, Im}
\newcommand {\rel} {\, Re}

\newcommand*\patchAmsMathEnvironmentForLineno[1]{%
\expandafter\let\csname old#1\expandafter\endcsname\csname #1\endcsname
\expandafter\let\csname oldend#1\expandafter\endcsname\csname
end#1\endcsname
 \renewenvironment{#1}%
   {\linenomath\csname old#1\endcsname}%
   {\csname oldend#1\endcsname\endlinenomath}%
}
\newcommand*\patchBothAmsMathEnvironmentsForLineno[1]{%
  \patchAmsMathEnvironmentForLineno{#1}%
  \patchAmsMathEnvironmentForLineno{#1*}%
}

\AtBeginDocument{%
\patchBothAmsMathEnvironmentsForLineno{equation}%
\patchBothAmsMathEnvironmentsForLineno{align}%
\patchBothAmsMathEnvironmentsForLineno{flalign}%
\patchBothAmsMathEnvironmentsForLineno{alignat}%
\patchBothAmsMathEnvironmentsForLineno{gather}%
\patchBothAmsMathEnvironmentsForLineno{multline}%
\patchBothAmsMathEnvironmentsForLineno{eqnarray}%
}
\usepackage{longtable} 
\usepackage{subdepth}

\begin{document}

\author{Biplab Dey}
\affiliation{Institute of Particle Physics, Central China Normal University, Wuhan, Hubei, China}
\date{\today}
\title{Towards a complete angular analysis of the electroweak penguin decay \Ltophll}

\begin{abstract}
We investigate the rare electroweak penguin transition $b \to s \ellp \ellm$ though angular analyses in baryonic $b$-decays. Instead of performing likelihood fits, that might suffer from ambiguity issues, we employ the moments technique, which is a simple counting experiment. The method is known to be robust under low statistics conditions and avoids the ambiguity problem by construction. We provide the full set of 32 orthonormal moments for \Ltophll, where $h\in\{\pi,K\}$, for a polarized $\Lb$ and the dihadron system $[ph]$ being in a spin-1/2 state.
\end{abstract}

\let\oldmaketitle\maketitle
\renewcommand\maketitle{{\bfseries\boldmath\oldmaketitle}}

\maketitle

\section{Introduction}

Among the dedicated $B$-factories, LHCb is unique in its ability to have access to all species of $b$-hadrons. Within the detector acceptance, the production rates of $B:\Lb:B_s$ are approximately in the ratio $4:2:1$~\cite{Aaij:2011jp}. The large number of $\Lb$ baryons available in a clean environment, thanks to the excellent vertexing and particle identification of the LHCb detector, has led to discoveries such as the Pentaquark states~\cite{Aaij:2015tga}, the first evidence of \CP violation in $b$-hadron decays~\cite{Aaij:2016cla}, among others.

In this work, we employ decays of the $\Lb$ baryon to study the rare electroweak penguin transition $b \to s \ellp \ellm$. In the Standard Model, this is a suppressed flavor-changing neutral current, occuring only via loop and box diagrams, and is therefore sensitive to new heavy particles entering the loop. In the mesonic sector, there has been some very interesting tensions with the Standard Model recently~\cite{Descotes-Genon:2015uva}, so it is important to probe this in as many modes as possible. LHCb has already made observations of the $\Lb \to \Lz \mup\mun$~\cite{Aaij:2015xza}, $\Lb\to p\Km\mup\mun$~\cite{Aaij:2017mib} and $\Lb\to p\pim\mup\mun$~\cite{Aaij:2017ewm} decays, using Run~1 data. Updates to these results, including Run~2 data, that will increase the statistical precision by a factor of four, are expected soon. The valence quark content of the $\Lb$ is $|bud\rangle$, where the $[ud]$ system often acts as a spectator diquark in a spin and isospin singlet. Therefore the properties of the $\Lb$ are expected to reflect those of the underlying $b$-quark. This is different from the $B$-meson states, where spin information of the $b$-quark is lost during the hadronization with the lighter quark to form a spin-0 meson. Therefore, the $\Lb$ mode can provide complementary information relative to the $B_{(s)}$ meson modes.

The LHCb tracking system~\cite{Alves:2008zz} consists of an innermost tracker (VELO) surrounding the interaction point, followed by two large-area trackers, one upstream (TT) and another downstream (tracking stations) of the LHCb magnet. Since the $\Lz$ flies considerably before decaying weakly, most of the $\Lz$ candidates decay outside the VELO (downstream tracks). In the low dimuon mass squared ($\qsq$) region, both the dimuon and dihadron system have sufficient break-up momenta and are more likely to be swept out of acceptance. Therefore, the $\Lz$ acceptance tends to be poor in the low $\qsq$ region. However, it is this low $\qsq$ (large-recoil) region that is most suited for theoretical predictions of observables with reduced uncertainties from hadronic effects. The high $\qsq$ (low-recoil) region receives contribution from broad $c\bar{c}$ resonances and is not theoretically clean. 

In this respect, the excited $\Lz^\ast\to pK$ and $N^\ast \to p\pi$ states can be particularly advantageous. These decays occur promptly inside the VELO and is relatively impervious to the low acceptance problem at high recoil. Given than angular observables have often proven to be sensitive to New Physics (NP) contributions~\cite{Descotes-Genon:2013wba}, it is desirable to measure as many angular observables for the $\Ltophll$ ($h \in \{ K,\pi\}$) cases as well. The goal of this paper is to set up the formalism to measure these observables employing the so-called moments technique, as outlined in Refs.~\cite{Dey:2015rqa,Dey:2016oun} for the $B$-meson case.

\section{The angle conventions}

\begin{figure}
 \centering
\subfigure[]{
\centering
 \includegraphics[width=0.45\textwidth]{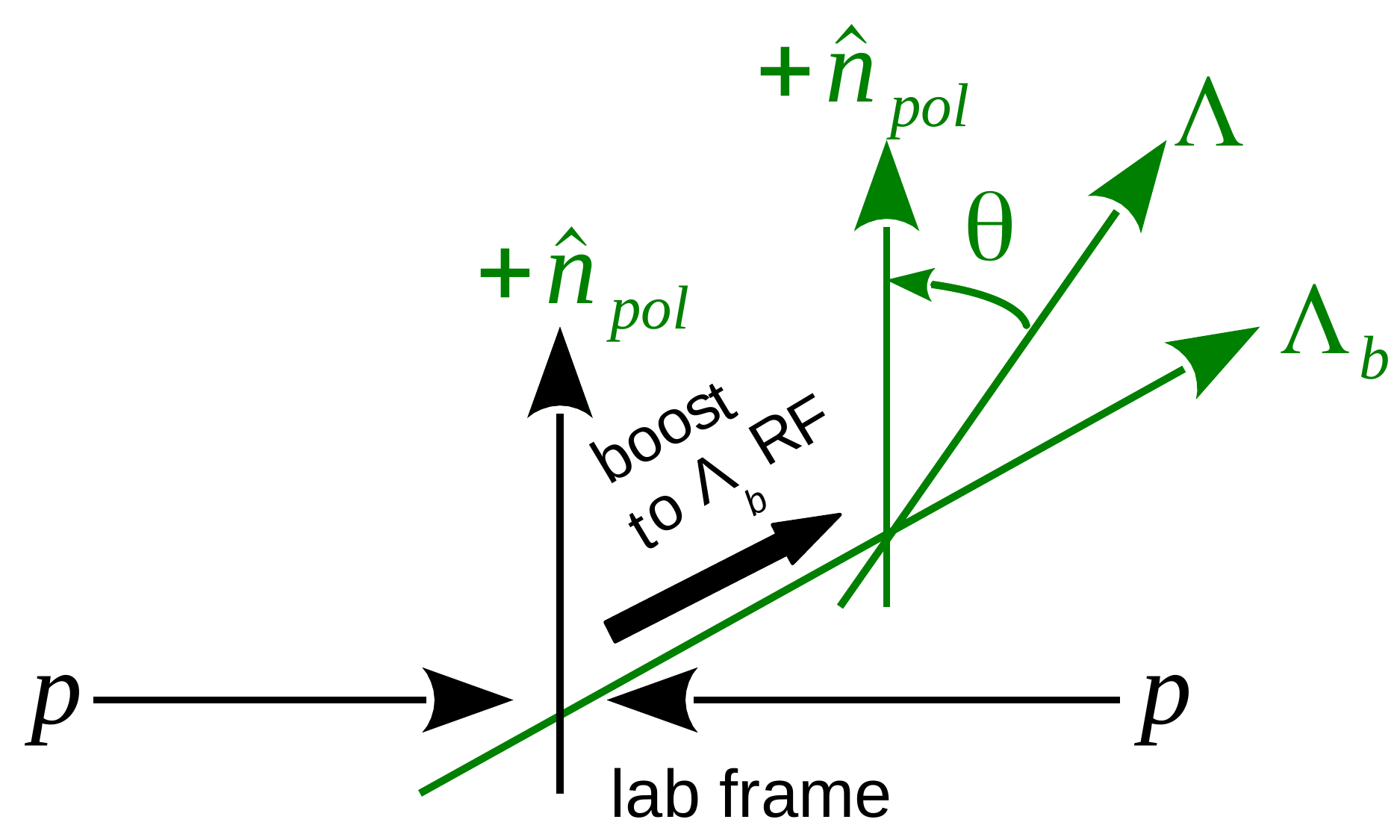}
}
\subfigure[]{
\centering
 \includegraphics[width=0.45\textwidth]{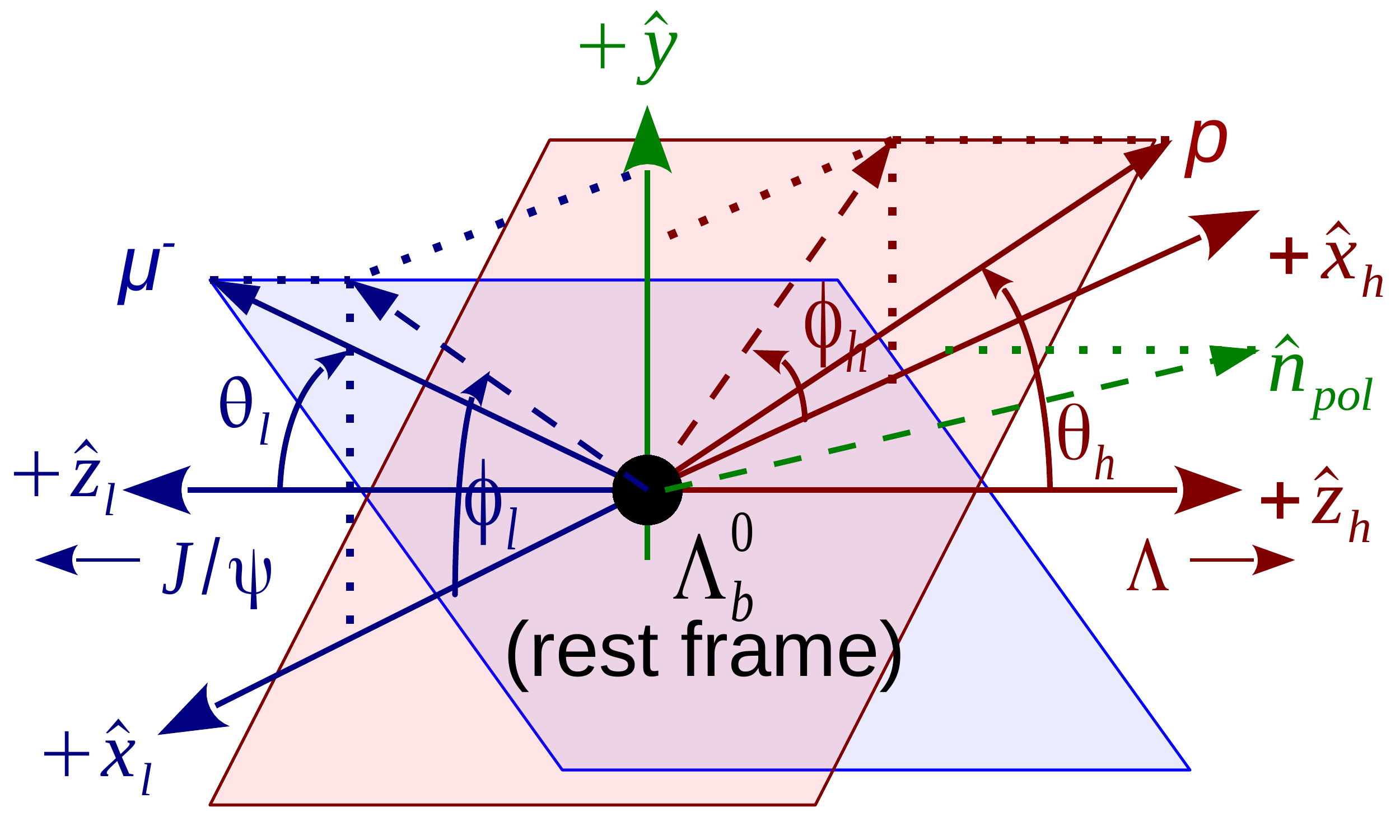}
}
\subfigure[]{
\centering
 \includegraphics[width=0.45\textwidth]{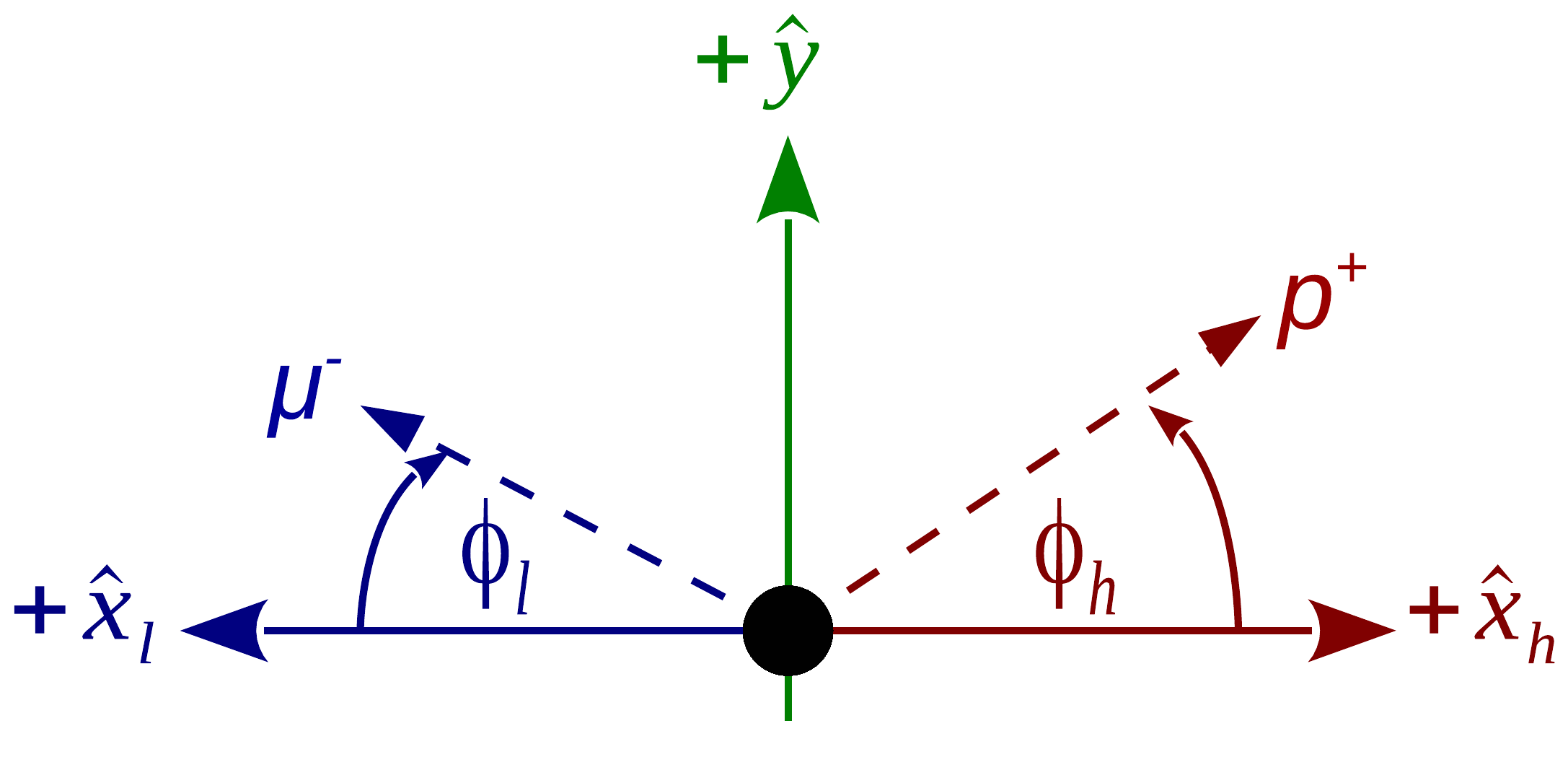}
}
 \caption{(color online) Angular conventions following Korner~\cite{Korner:2014bca}, shown for the specific case of the $\LJLzp$ decay. In (b), the $\Lb$ polarization vector $\hat{n}_{\rm pol}$ lies in the $\hat{x}_h$-$\hat{z}_h$ plane with a positive $\hat{x}_h$ intercept; the leptonic and hadronic decay planes are also shown. The leptonic and hadronic reference frames are back-to-back and share a common $\hat{y}$-axis. The orientations for the azimuthal angles are shown in again in (c). \label{fig:angle_defn}}
\end{figure}

We follow the conventions in Korner~\cite{Korner:2014bca} and Singleton~\cite{Singleton:1990ye} for $\Lb$ angular analyses. A pictorial representation of the angles for the specific case of the $\LJLzp$ decay is given in Fig.~\ref{fig:angle_defn}. In the laboratory frame, the $\Lb$ transverse polarization is along the normal to the plane formed by the $pp$ beam direction and the $\Lb$ flight direction. This normal direction is shown as $\hat{n}_{\rm pol}$ in Fig.~\ref{fig:angle_defn}a. We then boost to the $\Lb$ rest frame and measure the polar angle $\theta$, of the $\Lz$ with $\hat{n}_{\rm pol}$ in the $\Lb$ rest frame.

Next, we rotate the system such that the $\Lz$ flight direction points along the $+\hat{z}_h$ direction and $\hat{y}_h$ is along $\hat{z}_h\times \hat{n}_{\rm pol}$. That is, $\hat{n}_{\rm pol}$ lies in the $\hat{x}_h$-$\hat{z}_h$ plane with a positive intercept along the $\hat{x}$ direction. This fixes the quadrant of the azimuthal angles. The hadronic-side axes are labelled by the subscript $h$ here, and also in Fig.~\ref{fig:angle_defn}b.

The leptonic-side coordinate axes are defined in Fig.~\ref{fig:angle_defn}b and are labeled by the subscript $\ell$. The leptonic and hadronic systems have a common $\hat{y}$ axis and back-to-back $\hat{x}$ and $\hat{z}$ axes, conforming with the conventions in Korner~\cite{Korner:2014bca} and Singleton~\cite{Singleton:1990ye}. Figure~\ref{fig:angle_defn}c illustrates the signs of the azimuthal angles $\phi_{h,\ell}$.

We note here that compared to the ATLAS definitions~\cite{Aad:2014iba,Hrivnac:1994jx}, our azimuthal angles are shifted as $\phi_{h,\ell} \to \phi_{h,\ell} +\pi$.

\subsection{The decay amplitudes for \Ltophll}
\label{sec:dec-amp}


\begin{table}
\centering
\begin{tabular}{c|cc|c}
Amplitude $H^{J,\eta}_\lambda$ & $\lambda_{\Lz^\ast}/2$ & $\lambda_\psi/2$ & $\lambda/2 \equiv (\lambda_{\Lz^\ast} - \lambda_\psi)/2$ \\ \hline \hline
$C^{J,\eta}_+$ & $+3/2$ & $+1$ &  $+1/2$\\
$A^{J,\eta}_+$ & $+1/2$ & $\phantom{+}0$ &  $+1/2$\\
$B^{J,\eta}_+$ & $-1/2$ & $-1$ &  $+1/2$ \\ \hline
$C^{J,\eta}_-$ & $-3/2$ & $-1$ &  $-1/2$\\
$A^{J,\eta}_-$ & $-1/2$ & $\phantom{+}0$ &  $-1/2$\\
$B^{J,\eta}_-$ & $+1/2$ & $+1$ &  $-1/2$ \\ \hline
\end{tabular}
\caption{The twelve helicity amplitudes for the polarized decay \Ltophll. The helicities of the $\Lb$, spin-$J$ dihadron and spin-1 dilepton are denotes as $\lambda/2$, $\lambda_{\Lz^\ast}/2$ and $\lambda_\psi/2$, respectively. All amplitudes carry an additional $\eta=\pm 1$ tag for the leptonic-side .city The nomenclature follows Lednicky~\cite{Hrivnac:1994jx}.}
\label{tab:hel_amps}
\end{table}

Following Lednicky~\cite{Hrivnac:1994jx}, for the decay chain {\LJLzp}, we choose the spin-quantization axis to be along the $+\hat{z}_h$ direction, that is, the $\Lz$ flight direction in the $\Lb$ rest frame. This is shown in Fig.~\ref{fig:angle_defn}b. The topology is easily generalized to \Ltophll, with $h\in\{\pi,K\}$ and the dihadron system ($\Lz^{(\ast)}$, or $N^\ast$) being in a spin-$J$ state. For spin $J>1/2$, there are twelve amplitudes $\{A^{J,\eta}_\pm, B^{J,\eta}_\pm, C^{J,\eta}_\pm\}$ corresponding to the different helicity configurations of the dihadron and the dilepton systems as listed in Table~\ref{tab:hel_amps}. For spin $J=1/2$ states, only the $A^\eta_\pm$ and $B^\eta_\pm$ amplitudes contribute, as in the case of the ground state $\Lz$.

Following Korner~\cite{Korner:2014bca}, the full helicity amplitude can be concisely written in terms of the Wigner $d$-matrices as:
\begin{align}
H^{\lambda_p \eta}_\lambda&= \displaystyle \sum_J \Bigg[ \hspace{0.4cm} d^1_{\lambda \eta}(\thetal)\,\, C^{J,\eta}_\lambda \;\; d^J_{\frac{3 \lambda}{2} \frac{\lambda_p}{2}}(\thetah) e^{i\lambda\left(\frac{3\phih}{2} + \phil\right)} \nonumber \\
& \hspace{1cm} \;\;+ d^1_{0\eta}(\thetal) \;\; A^{J,\eta}_\lambda\;\; d^J_{\frac{\lambda}{2} \frac{\lambda_p}{2}}(\thetah) e^{i\lambda \frac{\phih}{2}} \nonumber \\
& \hspace{1cm} \;\;+ d^1_{-\lambda \eta}(\thetal) B^{J,\eta}_\lambda\, d^J_{-\frac{\lambda}{2} \frac{\lambda_p}{2}}(\thetah) e^{\mp i\lambda \left( \frac{\phih}{2} + \phil\right)} \Bigg] \nonumber \\
& \hspace{4cm} \times h^J_{\frac{\lambda_p}{2}},
\end{align}
where $\lambda_p/2$ is the helicity of the outgoing proton and $h^J_{\frac{\lambda_p}{2}}$ is the amplitude for the $\Lz^\ast \to ph$ decay. Here, $\eta = \lambda_{\mun} - \lambda_{\mup} = \pm 1$ and correspond to the handedness of the leptonic current. For left(right)-handed amplitudes, $\eta = -1$($+1$), while for the parity conserving electromagnetic transition $\jpsi \to \mun \mup$, the left- and right-handed amplitudes are the equal.

If only a single spin-$J$ component contributes, it is useful to define the asymmetry parameter
\begin{align}
\alpha = |h^J_+|^2 -  |h^J_-|^2
\end{align}
with the normalization set as
\begin{align}
1 = |h^J_+|^2 +  |h^J_-|^2
\end{align}
If the $\Lz^\ast$ state has a definite parity $P$, and decays strongly, then
\begin{align}
 h^J_{-\frac{\lambda_p}{2}}= (-1)^{J+1/2} P h^J_{\frac{\lambda_p}{2}},
\end{align}
resulting in $\alpha=0$. The weakly decaying $\Lz\to p \pim$ effectively includes both odd and even parity components, resulting in a non-zero $\allam$. 

\section{The differential rate}
\label{sec:diff-rate}

The differential rate expression is given by
\begin{align}
\label{eqn:diff_rate}
\frac{d\Gamma}{d\qsq d\Omega} &= (4\pi \sqrt{2}) \sum_{ \lambda_p, \eta, \lambda, \lambda' }  H^{\lambda_p \eta}_\lambda H^{\ast \lambda_p \eta}_{\lambda'} \rho_{\lambda \lambda'} \nonumber \\
                        &= (4\pi \sqrt{2})\;\;\; \sum_i \;\;\;\Gamma_i(\qsq) f_i(\Omega),
\end{align}
where $\qsq$ is the invariant dilepton mass squared, the phase-space element is 
\begin{align}
d\Omega \equiv d\cos\theta \;d\cth \;d\ctl \;d\phih \;d\phil,
\end{align}
 and the spin summation is over $\{\lambda_p, \eta, \lambda, \lambda'\} \in \pm 1$. The density matrix $\rho$ is given by
\begin{align}
\rho &= \frac{1}{2} (I + P_b \; \hat{n}_{\rm pol}\cdot \vec{\sigma})  \equiv \displaystyle \frac{1}{2} \begin{pmatrix} 1+ \pbee \cos\theta & \pbee \sin\theta \\ \pbee \sin\theta  & 1- \pbee \cos\theta\end{pmatrix},
\end{align}
where $ \vec{\sigma}$ are the spin-1/2 Pauli matrices. The $f_i$'s in Eq.~\ref{eqn:diff_rate} are a set of orthonormal basis functions and the $\Gamma_i$'s are the corresponding moments observables
\begin{align}
\Gamma_i(\qsq) \equiv \Gamma^L_i(\qsq) + \eta_i^{L\to R} \Gamma_i^R(\qsq),
\end{align}
where the sign $\eta_i^{L\to R}= \pm 1$ depends on the signature of $f_i(\Omega)$ under $\thetal \to \pi + \thetal$.

\subsection{Spin-1/2 case}

The expansion of Eq.~\ref{eqn:diff_rate} for dihadron spin-1/2-only case comprises 32 angular moments, as listed in Tables~\ref{tab:orth_mom} and~\ref{tab:orth_mom2}, including the two sets of $\eta_i^{L\to R}$ signs.  Here $\dzz \equiv \frac{1}{2} ( 3 \cos^2 \thetal - 1)$. For the parity conserving $\jpsi \to \ellm \ellp$ decay in \LJLz, only the first 20 terms with  $\eta_i^{L\to R}= + 1$ in Table~\ref{tab:orth_mom} contribute and our results are consistent with those listed in Ref.~\cite{Hrivnac:1994jx}. For the electroweak penguin case, 12 additional terms occur, with  $\eta_i^{L\to R}= - 1$. These are listed in Table~\ref{tab:orth_mom2}.

\begin{table*}
\centering
\resizebox{16cm}{!}{
\begin{tabular}{c|c|c|c}
 $i$ &   $f_i(\Omega) \times (4 \pi \sqrt{2})$             & $\Gamma_i^L$  & $\eta_i^{L\to R}$ \\ \hline \hline
 0 & $1$     &  $ \big(\apsq + \amsq + \bpsq + \bmsq \big)$ & +\\  \hline
 1 & $\sqrt{3}\cos\theta$     &  $ \displaystyle \frac{1}{\sqrt{3}}\left(\apsq - \amsq + \bpsq - \bmsq \right)$\pbee & "\\  \hline
 2 & $ \sqrt{3} \cos\thetah$     &  $ \displaystyle \frac{1}{\sqrt{3}}  \left(\apsq - \amsq - \bpsq + \bmsq \right) \alpha$ & "\\  \hline
 3 & $ 3 \cos\theta \cos\thetah$     &  $ \displaystyle \frac{1}{3} \left(\apsq + \amsq - \bpsq - \bmsq \right) $\pbee $\alpha$ & "\\  \hline
 4 & $\displaystyle \sqrt{5} \dzz$     &  $ \displaystyle \frac{1}{\sqrt{5}} \left(-\apsq - \amsq + \frac{\bpsq + \bmsq}{2}\right) $ & "\\  \hline
 5 & $\displaystyle \sqrt{15}\cos\theta \dzz$     &  $\displaystyle \frac{1}{\sqrt{15}}  \left(-\apsq + \amsq + \frac{\bpsq - \bmsq}{2}\right) $ \pbee& "\\  \hline
 6 & $\displaystyle \sqrt{15} \cos\thetah  \dzz$  &  $\displaystyle \frac{1}{\sqrt{15}}  \left(-\apsq + \amsq + \frac{-\bpsq + \bmsq}{2}\right)$$\alpha$ & "\\  \hline
 7 & $\displaystyle \sqrt{45} \cos\theta \cos\thetah \dzz$    &  $\displaystyle -\frac{1}{\sqrt{45}} \left(\apsq + \amsq + \frac{\bpsq + \bmsq}{2}\right)$ \pbee $\alpha$& "\\  \hline
 8 & $\displaystyle \frac{\sqrt{135}}{4}\sin\theta \sin\thetah \sin^2 \thetal \cos \phih $     &  $  \displaystyle \frac{4}{\sqrt{15}} \rel(A^L_+ A^{L\ast}_-)$ \pbee $\alpha$& "\\  \hline
 9 & $\displaystyle \frac{\sqrt{135}}{4}\sin\theta \sin\thetah \sin^2 \thetal \sin \phih $     &  $ - \displaystyle \frac{4}{\sqrt{15}} \img(A^L_+ A^{L\ast}_-) $ \pbee $\alpha$& "\\  \hline
 10 & $\displaystyle \frac{\sqrt{135}}{4} \sin\theta \sin\thetah \sin^2 \thetal  \cos(\phih + 2 \phil)$ &  $  \displaystyle  \frac{2}{\sqrt{15}} \rel(B^L_+ B^{L\ast}_-)$ \pbee $\alpha$& "\\  \hline
 11 & $\displaystyle \frac{\sqrt{135}}{4} \sin\theta \sin\thetah \sin^2 \thetal \sin(\phih + 2 \phil)$ &  $ \displaystyle  \frac{2}{\sqrt{15}} \img(B^L_+ B^{L\ast}_-)$ \pbee $\alpha$ & "\\  \hline
 12 & $\displaystyle \sqrt{\frac{135}{2}}\sin\theta \cos\thetah \sin\thetal \cos \thetal \cos \phil$ &  $ \displaystyle \sqrt{\frac{1}{15}} \rel( A^L_+ B^{L\ast}_- + A^{L\ast}_- B^L_+ )  $ \pbee $\alpha$& "\\  \hline
 13 & $\displaystyle \sqrt{\frac{135}{2}}\sin\theta \cos\thetah \sin\thetal \cos \thetal \sin \phil$ &  $ \displaystyle \sqrt{\frac{1}{15}} \img( A^L_+ B^{L\ast}_- + A^{L\ast}_- B^L_+)$ \pbee $\alpha$ & "\\  \hline
 14 & $\displaystyle \sqrt{\frac{135}{2}}\cos\theta \sin\thetah \sin\thetal \cos \thetal \cos(\phih+\phil) $ &  $ -\displaystyle \sqrt{\frac{1}{15}} \rel(A^L_+ B^{L\ast}_+ + A^L_- B^{L\ast}_-) $ \pbee $\alpha$ & "\\  \hline
 15 & $\displaystyle\sqrt{\frac{135}{2}}\cos\theta \sin\thetah \sin\thetal \cos \thetal \sin(\phih+\phil)$ &  $ \displaystyle \sqrt{\frac{1}{15}}  \img(A^L_+ B^{L\ast}_+ - A^L_- B^{L\ast}_-) $ \pbee $\alpha$ & "\\  \hline
 16 & $\displaystyle \sqrt{\frac{45}{2}}\sin\theta \sin\thetal \cos \thetal \cos \phil$ & $ \displaystyle -\frac{1}{\sqrt{5}}\rel(-A^L_+ B^{L\ast}_- + A^{L\ast}_- B^L_+) $ \pbee & "\\  \hline
 17 & $\displaystyle \sqrt{\frac{45}{2}} \sin\theta \sin\thetal \cos \thetal \sin \phil$ & $ \displaystyle  \frac{1}{\sqrt{5}} \img(A^L_+ B^{L\ast}_- - A^{L\ast}_- B^L_+) $  \pbee & "\\  \hline
 18 & $\displaystyle \sqrt{\frac{45}{2}}\sin\thetah \sin\thetal \cos \thetal \cos(\phih+\phil)$ & $ \displaystyle - \frac{1}{\sqrt{5}}  \rel(A^L_+ B^{L\ast}_+ - A^L_- B^{L\ast}_-) $ $\alpha$  & "\\  \hline
19 & $\displaystyle \sqrt{\frac{45}{2}}\sin\thetah \sin\thetal \cos \thetal \sin(\phih+\phil) $ & $ \displaystyle \frac{1}{\sqrt{5}} \img(A^L_+ B^{L\ast}_+ + A^L_- B^{L\ast}_-)\alpha$ & "
\end{tabular}
}
\caption{The moments of the first 20 orthonormal angular functions $f_i(\Omega)$ in Eq.~\ref{eqn:diff_rate} for the spin-1/2 case.}
\label{tab:orth_mom}
\end{table*}

\begin{table*}
\centering
\resizebox{13cm}{!}{
\begin{tabular}{c|c|c|c}
 $i$ & $f_i(\Omega) \times (4 \pi \sqrt{2})$             & $\Gamma_i^L$  &  $\eta_i^{L\to R}$ \\ \hline \hline
 20 &   $\displaystyle \frac{3}{\sqrt{2}} \sin\theta \sin \thetal \cos \phil$     & $ \rel (A^L_+ B^{L\ast}_- + A^{L\ast}_- B^L_+) \pbee$ & - \\  \hline
 21 &   $\displaystyle \frac{3}{\sqrt{2}} \sin\theta \sin \thetal \sin \phil$     & $ \img (A^L_+ B^{L\ast}_- + A^{L\ast}_- B^L_+) \pbee $  & " \\  \hline
 22 &   $\displaystyle 3 \sqrt{\frac{3}{2}}  \sin\theta \sin \thetal \cos \thetah \cos \phil$     &  $ \displaystyle \frac{1}{\sqrt{3}} \rel (A^L_+ B^{L\ast}_- - A^{L\ast}_- B^L_+)  \pbee \alpha    $ & " \\  \hline
 23 &   $\displaystyle 3 \sqrt{\frac{3}{2}}  \sin\theta \sin \thetal \cos \thetah \sin \phil$     & $ \displaystyle \frac{1}{\sqrt{3}} \img (A^L_+ B^{L\ast}_- - A^{L\ast}_- B^L_+)  \pbee \alpha  $  & " \\  \hline
 24 &   $\displaystyle \sqrt{3} \cos\thetal$     & $ - \displaystyle \frac{\sqrt{3}}{2} (\bpsq - \bmsq)$ & " \\  \hline
 25 &   $\displaystyle 3 \cos \thetah \cos \thetal $    & $ \displaystyle \frac{1}{2} (\bpsq + \bmsq)  \alpha $ & " \\  \hline
 26 &   $\displaystyle 3 \cos \theta \cos \thetal $     & $ - \displaystyle \frac{1}{2} (\bpsq + \bmsq  ) \pbee $ & " \\  \hline
 27 &   $\displaystyle 3 \sqrt{3}  \cos \thetah \cos \thetal \cos \theta$     & $ \displaystyle \frac{1}{2 \sqrt{3}} (\bpsq - \bmsq) \pbee \alpha   $ & " \\  \hline
 28 &   $\displaystyle \frac{3}{\sqrt{2}} \sin \thetal \sin \thetah \cos(\phih + \phil)$     & $ \rel(A^L_+ B^{L\ast}_+ + A^L_- B^{L\ast}_-)  \alpha  $ & " \\  \hline
 29 &   $\displaystyle \frac{3}{\sqrt{2}} \sin \thetal \sin \thetah \sin(\phih + \phil)$     & $ - \img(A^L_+ B^{L\ast}_+ - A^L_- B^{L\ast}_-) \alpha  $ & " \\  \hline
 30 &   $\displaystyle 3 \sqrt{\frac{3}{2}} \sin \thetal \sin \thetah \cos(\phih + \phil) \cos \theta$     & $ \displaystyle \frac{1}{\sqrt{3}} \rel(A^L_+ B^{L\ast}_+ - A^L_- B^{L\ast}_-)   \pbee \alpha  $ & " \\  \hline
 31 &   $\displaystyle 3 \sqrt{\frac{3}{2}} \sin \thetal \sin \thetah \sin(\phih + \phil) \cos \theta $     & $ - \displaystyle \frac{1}{\sqrt{3}}  \img(A^L_+ B^{L\ast}_+ + A^L_- B^{L\ast}_-) \pbee \alpha  $ & "
\end{tabular}
}
\caption{The 12 additional terms for the spin-1/2 dihadron, for the electroweak penguin case in Eq.~\ref{eqn:diff_rate}.}
\label{tab:orth_mom2}
\end{table*}

\subsection{\CP conjugation}
\label{sec:cp_conj}

The charge conjugation is performed ``explicitly''. With reference to Fig.~\ref{fig:angle_defn}, for the \CP conjugated $\Lbbar$ decay, we ``follow'' the $\mup$ and the $\bar{p}$. In the absence of direct \CP violation, this leads to the same form of the angular expression for $\Lb$ and $\Lbbar$ in terms of the corresponding helicity amplitudes. The helicity amplitudes between \CP conjugates are however related by
\begin{align}
H^{\lambda_p,\eta}_{\lambda}(\delta_W,\delta_S) = \overline{H}^{-\lambda_p,-\eta}_{-\lambda}(-\delta_W,\delta_S),
\end{align}
where $\delta_W$ and $\delta_S$ are the weak and strong phases. The polarization \pbee and the asymmetry $\alpha$ also flip sign. As can be checked from Tables~\ref{tab:orth_mom} and~\ref{tab:orth_mom2}, this effectively means that the signs of the measured moments odd in $\phi_{\ell,h}$ are flipped between \CP conjugates. To keep the signs of the measured moments the same between the \CP conjugates, we further flip $\phi_{\ell,h} \to -\phi_{\ell,h}$ for the $\Lbbar$. This is convenient to experimentalists since \CP symmtries can be measured by combining the $\Lb$ and $\Lbbar$ data samples.

It is important to note that the moments and amplitudes that we provide are given for the \Lb (containing a $b$-quark), since this is consistent with the conventions in several semileptonic and electroweak penguin theory papers.

\section{Ambiguities in the amplitude solutions}

Expanding out Eq.~\ref{eqn:diff_rate}, the differential rate is
\begin{align}
\frac{dN}{d\Omega} \;\sim &  \phantom{+}\;\; H^{+\eta}_+ H^{\ast +\eta}_+ \rho_{++}  + H^{+\eta}_- H^{\ast +\eta}_- \rho_{--} \nonumber \\ 
& \hspace{3cm} + 2 Re( H^{+ \eta}_+ H^{\ast + \eta}_-) \rho_{+-} \nonumber \\
&  +\; H^{-\eta}_+ H^{\ast -\eta}_+ \rho_{++}  + H^{-\eta}_- H^{\ast -\eta}_- \rho_{--} \nonumber \\ 
& \hspace{3cm} + 2 Re( H^{- \eta}_+ H^{\ast - \eta}_-) \rho_{-+}.
\end{align}
The above form is invariant under the the following set of simultaneous transformations: $\pbee\to -\pbee$ and $H^{\lambda_p \eta}_{\lambda} \to \lambda H^{-\lambda_p \eta}_{-\lambda}$. The $H^{\lambda_p \eta}_{\lambda} \to \lambda H^{-\lambda_p \eta}_{-\lambda}$ transformation can be affected by:
\begin{subequations}
\label{eqn:ambiguity_flips}
\begin{align}
C^{J,\eta}_\lambda &\to - i C^{\ast J,\eta}_{-\lambda} \\
A^{J,\eta}_\lambda &\to \phantom{-} i A^{\ast J,\eta}_{-\lambda}\\
B^{J,\eta}_\lambda &\to - i B^{\ast J,\eta}_{-\lambda} \\
h^J_{\lambda_p} &\to -i h^{\ast J}_{-\lambda_p}.
\end{align}
\end{subequations}
In particular, the flip in Eq.~\ref{eqn:ambiguity_flips}d means that the sign of $\alpha$ will also flip. The simultaneous flip of $\alpha$ and $\pbee$ is discussed in more detail in the subsequent section. In the case of $\pbee=0$, further ambiguities are also possible. 

The underlying reason for these ambiguities is that, as in the mesonic cases~\cite{Hofer:2015kka}, the spins of the final state leptons are not measured, but averaged over. Further, if the spin of the proton is also not measured (as for the $\Lz^\ast$ and $N^\ast$ cases), this leads to further ambiguities. The ambiguities imply that the full set of $\Gamma_i$ elements are not independent, but relations exist among them~\cite{Hofer:2015kka}. For an observables fit, it is necessary to identify these relations and the minimal set of independent observables that can be the variable parameters in the fit. The novelty of the moments technique is that ambiguity issue is rendered irrelevant since no fit is performed. Once the full set of moments (which might not be independent) and the covariance matrix is extracted from the data, these can be compared against model predictions.

\section{Solving for the \LJLzp amplitudes}
\label{sec:lb2jpsil0_solns}

Since the basis in Eq.~\ref{eqn:diff_rate} is constructed out of an orthonormal set of angular functions, the moments observables can be extracted out by weighting the data events by the angular functions themselves:
\begin{align}
\Gamma_i = \displaystyle \sum_{k \in N_{\rm data}} f_i(\Omega_k) / (4 \pi \sqrt{2}).
\end{align}
The full description of the formalism to perform background subtraction and detector acceptance correction was provided in Ref.~\cite{Dey:2015rqa} and was successfully used to perform an angular analysis of the $\Bzb\to \Km \pip \mun \mup$ system using the full Run~I LHCb dataset in Ref.~\cite{Aaij:2016kqt}.

For the $\jpsi \to \mun \mup$ mode, only the twenty moments in Table~\ref{tab:orth_mom} contribute. Further, the overall normalization is set to unity by switching to the set of normalized moments $\overline{\Gamma}_i \equiv \Gamma_i/\Gamma_0$. Denoting $A_\pm \equiv a_\pm e^{i\alpha_\pm}$ and $B_\pm \equiv b_\pm e^{i\beta_\pm}$, and $\allam$ as the weak-decay asymmetry, the magnitudes are solved as:
\begin{subequations}
\begin{align}
a_\pm^2 &= \displaystyle \phantom{-}\frac{1}{4 \pbee \allam} \left[(\pbee \allam + 3 \overline{\Gamma}_3) \pm \sqrt{3} (\allam \overline{\Gamma}_1 + \pbee \overline{\Gamma}_2)\right] \\ 
b_\pm^2 &= \displaystyle \phantom{-}\frac{1}{4 \pbee \allam} \left[(\pbee \allam - 3 \overline{\Gamma}_3) \pm \sqrt{3} (\allam \overline{\Gamma}_1 - \pbee \overline{\Gamma}_2)\right],
\end{align}
\end{subequations}
and also
\begin{subequations}
\begin{align}
a_\pm^2 &= \displaystyle -\frac{\sqrt{5}}{4 \pbee \allam} \left[(\overline{\Gamma}_4 \pbee \allam + 3 \overline{\Gamma}_7) \pm \sqrt{3} (\allam \overline{\Gamma}_5 + \pbee \overline{\Gamma}_6)\right] \\
b_\pm^2 &= \displaystyle -\frac{\sqrt{5}}{2 \pbee \allam} \left[(\overline{\Gamma}_4 \pbee \allam - 3 \overline{\Gamma}_7) \pm \sqrt{3} (\allam \overline{\Gamma}_5 - \pbee \overline{\Gamma}_6)\right].
\end{align}
\end{subequations}
These two sets of solutions yield the relations:
\begin{subequations}
\begin{align}
\pbee \allam &= \displaystyle - \frac{3 (\sqrt{5} \overline{\Gamma}_7 + \overline{\Gamma}_3)}{(1 + \sqrt{5}\overline{\Gamma}_4)},  \\
\frac{\allam}{\pbee} &= \displaystyle - \frac{(\sqrt{5} \overline{\Gamma}_6 + \overline{\Gamma}_2)}{(\sqrt{5} \overline{\Gamma}_5 + \overline{\Gamma}_1)},  \\
\pbee \allam &= \displaystyle \phantom{-}\frac{3 ( \overline{\Gamma}_3 -2\sqrt{5} \overline{\Gamma}_7)}{(1 - 2 \sqrt{5}\overline{\Gamma}_4 ) }, \; {\rm and} \\
\frac{\allam}{\pbee} &= \displaystyle  \phantom{-}\frac{(\overline{\Gamma}_2 - 2 \sqrt{5} \overline{\Gamma}_6)}{(\overline{\Gamma}_1 - 2 \sqrt{5}\overline{\Gamma}_5)  },
\end{align}
\end{subequations}
from which both $\pbee^2$ and $\allam^2$ can be {\em solved} out. This is the essence of the moments technique, where no fit is required. The full set of the complex amplitudes $A_\pm$ and $B_\pm$ can be solved out from the moments, up to an overall sign ambiguity in $\pbee$ and $\allam$. Of course, if the sign of \allam is known from elsewhere, the ambiguity is broken and the sign of \pbee is also known.

\section{Outlook and further work for higher spins}

In the present work, we gave the prescription for performaing an angular analysis of $\Lb$ decays in a manner that bypasses any relevant ambiguity issue. The moments calculations presented hold for generic spin-1/2 $\Lambda^\ast$ or $N^\ast$ hadronic final states. The next step would comprise extending the calculations (using the same formalism as in Sec.~\ref{sec:dec-amp}) to spin-3/2 and 5/2 cases. The procedure is straightforward, but the explicit expansion and rearrangement of terms has to be performed; this is currently a work in progress and will be covered in a forthcoming paper. Existing quark model predictions~\cite{Mott:2015zma} for $\Lb \to \Lambda^\ast \ellp \ellm$ indicate that $J\leq5/2$ would suffice for $m(pK)\leq 1900$~MeV. Given that most of the $\Lambda^\ast$ statistics lie in the low $m(pK)$ region, the calculation would enable an angular analysis for $\Lb \to p\Km\mup \mun$ in the high recoil region at LHCb.

\ifthenelse{\boolean{wordcount}}{ 
  \bibliographystyle{plainat}
  \nobibliography{biblio}
}{
 \bibliography{biblio}
}

\end{document}